\documentstyle[psfig]{l-aa}

\begin{document}

\thesaurus{02 (12.12.1; 12.07.1; 11.01.2, 13.25.2, 13.25.3)} 

\title{The x-ray background-foreground galaxy cross-correlation:
evidence for weak lensing?} 

\author{Asantha R. Cooray}

\institute{Department of Astronomy and Astrophysics, University of
Chicago, Chicago IL 60637, USA. E-mail: asante@hyde.uchicago.edu}

\date{Received: June 9 1999 ; accepted: July 2 1999}
\maketitle

\begin{abstract}
A significant cross-correlation between
the unresolved X-ray background (XRB) at soft energies (0.5 to 2 keV)
and foreground bright galaxies has now been reported in several
studies.  This cross-correlation has been 
interpreted in terms of a low
redshift and a low luminosity active galactic nuclei (AGN) population,
clustered similar to optically
bright galaxies, as responsible for the unresolved component of the XRB.
In contrast to such a low redshift population, we suggest that a
correlation between the unresolved XRB and bright optical galaxies 
can exist due to a high redshift population of
X-ray emitting AGNs through weak lensing effects of
low redshift large scale structure traced by foreground optical
galaxies.  We further
investigate this possibility and suggest that a substantial 
fraction of the detected cross-correlation signal can arise from
this scenario. The most likely explanation for the observed 
cross-correlation is that both a population of low redshift 
sources and a population of high redshift low luminous 
sources contribute through clustering and lensing
effects, respectively. The exact weak gravitational lensing 
contribution to the detected signal can eventually be used to constrain 
cosmological parameters, foreground galaxy bias and, more
importantly, models of high 
redshift X-ray emitting sources.

\end{abstract}


\section{Introduction}

After many years of observational work and theoretical investigations,
the nature and origin of the unresolved component of the cosmic X-ray
background (XRB) still remains an unsolved problem.  The deep X-ray imaging
data, combined with optical spectroscopic observations, now suggest that
up to $\sim$ 70\% of the  soft XRB
observed with ROSAT in the 0.5 to 2.0 keV energy band
is resolved to individual galaxies, mainly active galactic nuclei (AGN),
 out to redshifts of $\sim$ 4 and greater (e.g., Miyaji et al. 1998a;
Hasinger  1999 contains a recent review).
Other than various possibilities that
have been suggested in the literature,
the exact nature of the remaining contributors to the soft XRB has
not been clearly established. 
The possibilities for candidates so far include a population of low-luminosity
galaxies and AGNs, and an optically obscured population of moderate to high
redshift and high luminosity galaxies and AGNs. The strong isotropy of
the unresolved component of the XRB, as measured by its
auto-correlation function, requires that most of the sources
responsible are at high redshifts and constraints models
involving a population of low redshift and low luminosity AGNs.
Returning to an obscured population at optical wavelengths,
the hard XRB requires a ratio of obscured to unobscured 
populations of AGNs that
amount to a factor as high as $\sim$ 3; As discussed in Almaini et
al. (1999), the implications for such an obscured
population is wide ranging.

Recent experimental developments now allow some of these
possibilities to be observationally tested.
For example, the obscured population at optical wavelengths is
expected to be visible at submm and far-infrared (FIR) wavelengths, through
reemission of absorbed UV radiation by dust at longer wavelengths. 
Such sources should now be
detected through deep observations with Submm Common User Bolometer
Camera (SCUBA; Holland et al. 1998) on the James Clerk Maxwell
Telescope. The current ongoing deep surveys with SCUBA will eventually
test the exact fraction of obscured AGNs (see, Smail et al. 1999 for
a recent review), with initial results suggesting that a dominant AGN 
fraction as high as 30\% may be contributing to current SCUBA number
counts (e.g., Cooray 1999a). At hard X-ray wavelengths, most of the 
Compton-thick AGNs which are  absent at soft X-ray bands 
are expected to be present.
Such populations have now been searched with ASCA and the
Italian-Dutch BeppoSax satellite (Piro et al. 1995)
in the 2 to 10 keV energy band. Contrary to expectations, however,
these surveys are finding that all hard X-ray sources have soft X-ray
counterparts (Hasinger 1999; however, see, Fiore et al. 1999). 
As most of these FIR/submm and hard
X-ray observational programs are
still ongoing, it is unlikely that an exact answer on the 
sources responsible for the unresolved component will soon be available.

Recently, the existence of a high redshift population of low luminous
X-ray emitting sources has been suggested
by Haiman \& Loeb (1999). These sources are present in cosmological
models of hierarchical structure formation and are associated with the
first generation of  quasars. The presence of a high redshift population of
X-ray emitting sources is also suggested by the possibility that there is
no clear evidence for a decline in X-ray AGN number counts
beyond a redshift of 2.5 (e.g., Miyaji et al. 1998b), which is contrary
to optical quasar surveys where a decline has been inferred at
high redshifts (e.g., Schmidt et al. 1995). According to the 
expected number counts of
high redshift AGNs from Haiman \& Loeb (1999),
the contribution to current unresolved XRB from a high redshift 
AGN population is greater than 90\%. Thus, almost all of the
present unresolved XRB can be explained with such a low X-ray luminous
population and without invoking the presence of
optically obscured or Compton-thick sources.
In addition to analytical calculations presented in Haiman \& Loeb
(1999), a population of high redshift low luminous quasars is
also present in Monte Carlo realizations of merger 
histories of dark matter halos based
on extended Press-Schechter theory (see, e.g. Cole 1991; Kauffmann \&
White 1993; Somerville \& Kolatt 1998) combined with
semi-analytical models of galaxy and quasar formations (Cooray \&
Haiman, in preparation). Given that the direct detection of such low
luminous AGNs  at X-ray wavelengths 
 is not likely to be possible with current observational programs, 
the evidence for such  high redshift X-ray sources should be inferred 
through indirect methods. It is likely that this situation will soon
change with upcoming X-ray satellites such as the Chandra X-ray
Observatory (CXO)\footnote{http://asc.harvard.edu} and
the X-ray Multiple Mirror (XMM) Telescope 
\footnote{http://astro.estec.esa.nl/XMM/}.

In Almaini et al. (1997), a cross-correlation between the unresolved
XRB at soft X-ray energies, based on three $\sim$ 50 ksec 
ROSAT deep wide-field deep	
observations, and foreground bright
galaxies, down to B-band magnitude of 23,
 has been presented. Such a correlation has been previously
investigated in various studies involving the nature of XRB and
sources responsible for it (e.g., Lahav et al. 1993; Miyaji et al.
1994; Carrera et al. 1995; Roche et al. 1996; Refregier et al. 1997;
Soltan et al. 1997).
The cross-correlation showed a highly 
significant signal and has been interpreted as
evidence for a population of low redshift sources, traced by bright
optical galaxies, as contributors to the unresolved XRB. 
Such an interpretation is based on the fact that detected cross-correlation is
due to clustering between sources responsible for the unresolved
component of  the XRB and optical galaxies. If clustering were not to
be present, in a case in which sources responsible for the XRB and
optical galaxies were physically distinct in redshift space - or at
least at scales greater than $\sim$ 100 Mpc - one would not normally
expect any cross-correlation signal to be present. Apart from
clustering, however, physically distinct populations can produce detectable
cross-correlation if the flux-limited number counts and/or spatial
distribution of one population was affected by the other.
A well known possibility is that gravitational lensing by foreground
sources  modifies the
distribution and number counts of background sources.
Thus, an alternative possibility for the unresolved XRB is a 
population of sources at high redshifts provided
that their X-ray emission  is gravitationally lensed through 
foreground large scale structure.
The cross-correlation between such sources and foreground optical
galaxies results from the fact that
foreground galaxies are a biased tracer of the
large scale structure. 
The presented cross-correlation
effect here is similar to the one involving high redshift optical
quasars and foreground galaxies as discussed in Bartelmann (1995)
and Dolag \& Bartelmann (1997). A more general treatment of the
cross-correlation between foreground and background samples due to
weak gravitational lensing could be found in Sanz et al. (1997) and
 Moessner \& Jain (1998). In both these studies, cross-correlation between two
distinct populations in redshift was suggested as a probe of weak lensing due 
to large scale structure.  In Sect. 2, we further 
investigate this possibility by  modeling the X-ray emission from 
background sources and considering 
weak lensing effects of X-ray number counts. 
We use recent results from Haiman
\& Loeb (1999) to describe the background X-ray population.
The general framework for the weak lensing calculation follows
Cooray (1999b). We refer the reader to
Mellier (1998) for a recent review on weak gravitational lensing,
its applications and observations.
We follow the conventions that the Hubble constant,
$H_0$, is 100\,$h$\ km~s$^{-1}$~Mpc$^{-1}$ and
 $\Omega_i$ is the fraction of the critical
density contributed by the $i$th energy component: $b$ baryons, $\nu$
neutrinos, $m$ all matter species (including baryons and neutrinos)
and $\Lambda$ cosmological constant.

\section{The x-ray background - foreground galaxy cross-correlation}

Here, we briefly describe the expected signal between a foreground
galaxy population with number density $n_g$ and 
X-ray sources responsible for the unresolved XRB, $n_x$. The angular
cross-correlation function between the two samples is:
\begin{equation}
w(\theta) = \langle \delta n_g (\hat{\phi}) \delta n_x (\hat{\phi}')\rangle
\end{equation}
where $\delta n$ is the excess fluctuations at a given line of sight.
The cross-correlation between two physically distinctive samples
contain four terms (Moessner \& Jain 1998):
\begin{eqnarray}
w(\theta) = & \langle \delta n_g^c (\hat{\phi}) \delta n_x^c
(\hat{\phi}')\rangle
+ \langle \delta n_g^c (\hat{\phi}) \delta n_x^\mu
(\hat{\phi}')\rangle
 \\ \nonumber
& + \langle \delta n_g^\mu (\hat{\phi}) \delta n_x^c (\hat{\phi}')\rangle
+ \langle \delta n_g^\mu (\hat{\phi}) \delta n_x^\mu (\hat{\phi}')\rangle,
\end{eqnarray}
where $\delta n^c$ is the fluctuations due to clustering of the sources
while $\delta n^\mu$ is fluctuations due to gravitational lensing.
These two terms can be written as,
\begin{equation}
\delta n^c(\hat{\phi}) = \int_0^{\chi_H} d\chi \,
b(r(\chi)\hat{\phi},\chi) 
W(\chi) \delta(r(\chi)\hat{\phi},\chi)
\end{equation}
and,
\begin{equation}
\delta n^\mu(\hat{\phi}) = 3 (\alpha - 1) \Omega_m
\int_0^{\chi_H} d\chi \, g(\chi) \delta(r(\chi)\hat{\phi},\chi),
\end{equation}
respectively. Here, 
$\chi_H$ is the comoving distance to the horizon,
$W(\chi)$ is the radial distribution of sources,
$\alpha$ is the slope of number counts of these sources, 
$n \propto S^{-\alpha}$ with flux $S$,
$b(r(\chi)\hat{\phi},\chi)$ is the source bias with respect to
matter distribution, assuming to be both scale and time dependent, and
$g(\chi)$ is a weight function:
\begin{equation}
g(\chi) = r(\chi) \int_{\chi}^{\chi_H}
\frac{r(\chi'-\chi)}{r(\chi')} W(\chi')
d\chi'.
\end{equation}
In Eq.~(3), (4) and (5),
$r(\chi)$ is the comoving angular diameter distance
written as $r(\chi) = 1/\sqrt{-K} \sin \sqrt{-K}\chi,\chi,
1/\sqrt{K}\sinh \sqrt{K}\chi$ for closed, flat and open models
respectively with $K = (1-\Omega_{\rm tot})H_0^2/c^2$
and $\chi$ is the radial comoving distance related to redshift
$z$ through:
\begin{equation}
\chi(z) =  \frac{c}{H_0}\int_{0}^{z} dz' \left[ \Omega_m (1+z')^3 +
\Omega_k (1+z')^2 + \Omega_\Lambda \right]^{-1/2}.
\end{equation}
The lensing term in the cross-correlation is due to the fact that
number counts of lensed background sources are affected in two ways:
magnification by a factor $\mu$ so that lensed counts reach a fainter
flux level ($S/\mu$) and distortion of the observed area such that
solid angle observed is reduced by a factor $1/\mu$. Thus, lensed
number counts change to $n' \propto \mu^{\alpha -1} S^{-\alpha}$ from
unlensed counts of $n \propto S^{-\alpha}$. In the weak lensing limit,
magnification $\mu = 1+ 2\kappa$, where $\kappa$ is the convergence
and is equivalent to a weighted projection, via $g(\chi)$, 
of the matter distribution
along the line of sight to background sources (see, e.g.,
Jain \& Seljak 1997; Kaiser 1998;  Schneider et al. 1998).

The four terms in the cross-correlation are 
respectively: (1) clustering of sources in the two
samples, when their redshift distributions overlap (2)
lensing of background sources by large scale structure front of them
traced by foreground galaxies
(3) lensing of foreground sources by large scale structure traced by
background galaxies; this term is
 non-zero only if there is an overlap in redshift
distribution between the two samples, and (4)
lensing of both foreground and background sources by large scale
structure. 

When there is no overlap in redshift between the two samples, terms
(1) and (3) are zero, while the last term can be ignored
as its contribution is an order of magnitude lower than the 2nd term
involving lensing of background sources by foreground large scale
structure. The gravitational lensing effect results from two effects:
(1) magnification due to lensing such that sources too faint to be
included due to flux limit are now introduced and (2) modification of
 the observed solid angle, or volume, such that number counts are diluted.
Considering these two well known effects, finally, the
cross-correlation between two samples separated in redshift space 
can be written in the weak lensing limit as:
\begin{eqnarray}
w_{gx}(\theta) = & 3 b_g \Omega_m (\alpha_x - 1) \int_{0}^{\chi_H}
W_g(\chi) \frac{g_x(\chi)}{a(\chi)} \\ \nonumber
& \int_{0}^{\infty} \frac{dk\, k}{2 \pi} P(k,\chi)
J_0\left[k r(\chi) \theta \right]
\end{eqnarray}
where
$W_g(\chi)$ and $W_x(\chi)$ are the radial distributions
of foreground galaxies and background X-ray sources,
$\alpha_x$ is the slope of number counts of background X-ray emitting  sources
at the limit of the unresolved background, and $b_g$ is the
galaxy bias, assuming a linear bias independent of scale and time.  
The detailed derivation of Eq.~(7), can be found in
Bartelmann (1995) for an Einstein-de Sitter Universe with an extension
to general cosmologies and nonlinear evolution of the power spectrum
in Dolag \& Bartelmann (1997), Sanz et al. (1997) and Moessner \& Jain (1998). Here, we have introduced slope of the number
counts, $\alpha_x$, for background X-ray sources, while, for example, 
in Moessner \& Jain (1998) background sources were considered to be galaxies
with a logarithmic number count slope of $s$ in magnitudes.

\subsection{Expected contribution from weak lensing}

In order to estimate the expected level of contribution from weak
lensing effects, we describe the background X-ray sources following
calculations presented in Haiman \& Loeb (1999). The foreground
sources are described following Almaini et al. (1997), with a redshift
distribution that peaks at a redshift of $\sim$ 0.5 and decreases to zero
by redshift around $\sim$ 2.0. Such a redshift 
distribution for galaxies down to a
magnitude limit of 23 in B-band is consistent with observations.
We assume that galaxies are biased such that $b_g=1/\sigma_8$, which 
should adequate for the present calculation.
Since most of the galaxies are at low redshifts, our predictions are
insensitive to the exact redshift distribution of background sources
as long as their redshifts are greater than 2.0.
For the purpose of this calculation, 
we consider a background redshift distribution 
in which X-ray sources are distributed around a mean redshift of
$\sim$ 3.5. In Fig.~1, we show the two foreground and
background redshift distributions. There is a slight overlap in
redshift between the two distributions, but we have ignored it for the
purpose of this calculation. 

\begin{figure}
\psfig{file=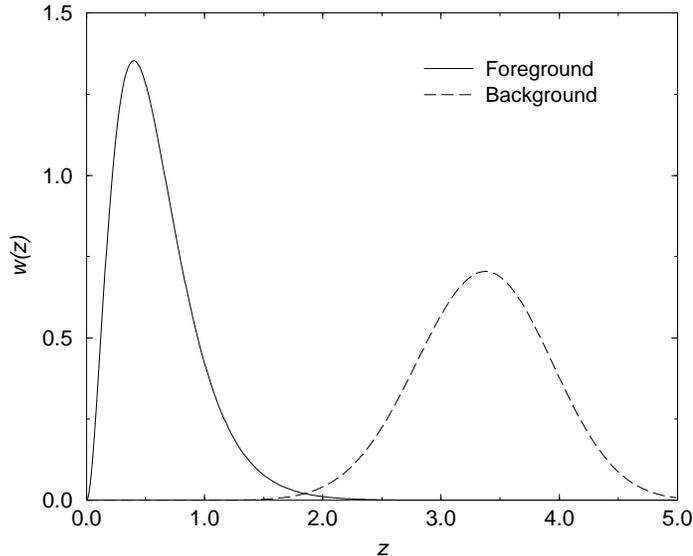,width=3.6in,angle=-90}
\caption{The redshift distribution of foreground galaxy and background
X-ray sources.}
\end{figure}

Our input dark matter power spectrum and its non-linear evolution is
calculated following Cooray (1999) using the fitting formulae
given in Hu \& Eisenstein (1998) to obtain the transfer function and
Peacock \& Dodds (1996) to obtain the nonlinear evolution. 
We consider cosmologies in which $\Omega_b =0.05$, $\Omega_\nu=0.0$, 
$h=0.65$. The power
spectrum is normalized to $\sigma_8 (= 0.56 \Omega_m^{-0.47})$ as 
determined by number density of galaxy clusters (Viana \& Liddle 1996).
Following calculations presented in Haiman \& Loeb (1999), we
determined the slope of X-ray number counts, $\alpha$, at the limit of the
unresolved XRB to be $\sim$ 1.2. This number, however, is not well
determined and is highly sensitive to how one models the X-ray
emission from high redshift low luminous sources and number counts
of such sources, as derived based on the Press-Schechter theory.
We note that a value for $\alpha < 1.0$ produces
a cross-correlation which is negative, while $\alpha=1.0$ produces
no contribution to cross-correlation from weak lensing.

Finally, in order to account for the finite point spread function (PSF) of
the PSPC detector, we convolve the expected lensing contribution with
a parametric form of the PSF given by Hasinger et al. (1992).
In Fig.~2, we show the expected contribution from weak lensing to be
observed cross-correlation. The data and associated errors are from
Almaini et al. (1997). The two curves show the expected contribution
for two cosmological models involving $\Omega_m=1.0$ and
$\Omega_m=0.3$, $\Omega_\Lambda=0.7$.

\begin{figure}
\psfig{file=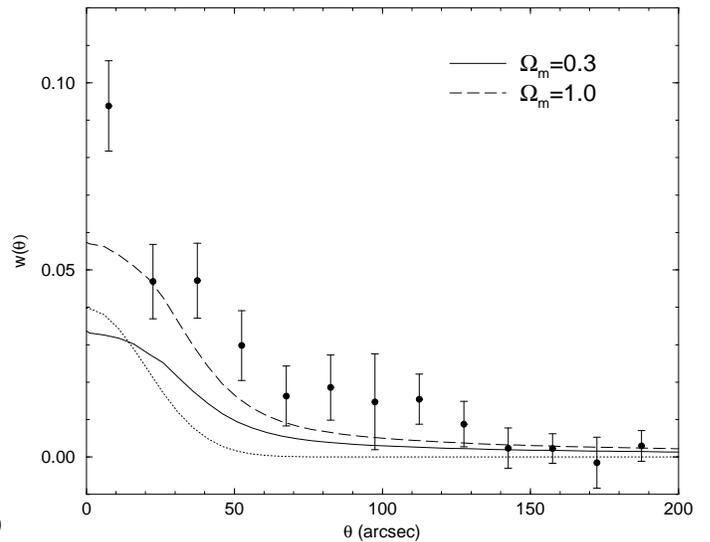,width=3.6in,angle=-90}
\caption{The observed cross-correlation between the XRB and foreground
galaxies. Data are from Almaini et al. (1997). The two curves show
the expected weak lensing contribution for two cosmological models
involving
$\Omega_m=1.0$ (dashed line) and $\Omega_m=0.3$, $\Omega_\Lambda=0.7$
(solid line). The dotted line is the expected Poisson noise
contribution to the cross-correlation (see text).}
\end{figure}

\section{Discussion \& summary}

As shown in Fig.~2, the weak lensing contribution to the observed
cross-correlation between the XRB and foreground bright
galaxies is substantial. The fractional contribution in the simple
model considered here amount up to and more than 50\%. 
Ignoring such a contribution is likely to produce
biased estimates on the amplitude of clustering
or the luminosity density
of X-ray sources. In this respect, we note that previous estimates on
the number density and luminosities of sources responsible for the
unresolved XRB, using the cross-correlation, is certainly overestimated.
In addition to changing cosmological parameters, we can increase
the lensing contribution by increasing the foreground galaxy bias or
increasing the slope of the X-ray number counts at the limit of the
unresolved XRB. Currently, both these 
quantities, more importantly the slope of the number counts, are
unknown. Therefore, it is premature to consider detailed models to
explain the XRB using weak lensing effects completely.
Since galaxy bias, however, 
is not expected to be much larger than $1/\sigma_8$,
especially at low redshifts considered here,  and that the
slope of number counts is not likely to be very steep, it is unlikely
that weak lensing alone can be used 
to fully explain the observed cross-correlation signal.

As shown in Fig.~2, weak gravitational lensing and Poisson
fluctuations can easily account for almost all of the detected
cross-correlation. However, we note that, in addition to weak lensing
by large scale structure, strong lensing by individual galaxies and
clusters of galaxies can contribute to the observed signal at small
lag angles. Such a contribution is likely to be smaller than the weak
lensing effect; still, it is likely that we have underestimated the
complete lensing contribution to cross-correlation between the unresolved
XRB and foreground galaxies by only considering weak lensing effects.

The most likely scenario is that the observed cross-correlation is
both due to clustering, from a low redshift population overlapping
with the galaxy distribution and weak lensing  effects
of a high redshift population. 
The isotropy of the XRB, from its 
auto-correlation function, requires that bulk of the sources are
at redshifts greater than 1. The clustering analysis of the observed
XRB-galaxy cross-correlation suggests that up to $\sim$ 40\% of the
unresolved XRB is due to faint low-redshift X-ray sources (e.g.,
Almaini et al. 1997; Roche et al. 1996; Soltan et al. 1997). The
additional contribution could arise from the high redshift X-ray
emitting sources, however, we note that intracluster medium of 
galaxy clusters and groups as well $\sim$ 10$^{6}$ Kelvin gas in 
outskirts of galaxies, where most of the baryons at low redshifts are 
now believed to be present (Cen \& Ostriker 1999), can contribute to
the unresolved XRB.

In addition to clustering and lensing terms, 
an additional term is 
present in the cross-correlation at zero lag or when $\theta=0$
due to the Poisson behavior of the background.
Even though this term only
arises for $\theta=0$, the finite PSF produces
a substantial contribution at angular separations out to
$\sim$ 30 arcsecs; the contribution is proportional to the
integrated luminosity density of X-ray sources. 
Following Almaini et al. (1997) and using the Miyaji et
al. (1998a)
luminosity function for X-ray AGNs at a redshift of $\sim$ 3.5,
we have estimated such a Poisson noise contribution to the
cross-correlation. In Fig.~2, we show this term with a dotted line.
A Poisson fluctuation contribution level similar to the one calculated
and a weak lensing contribution similar to the one calculated
for $\Omega_m=1.0$, when added, can easily explain the observed
cross-correlation signal. 
As stated earlier, given that we have no reliable knowledge
on the number counts and foreground galaxy bias, such a fit to the
observed data is meaningless.  We leave the task of a detailed
comparison between the observed XRB and galaxy cross-correlation 
and various models involving  lensing, clustering and Poisson
contributions to a later paper. In fact, if the contribution
to cross-correlation from latter two terms can be independently
determined, then the lensing contribution  can be used as a probe of
the high redshift low luminosity X-ray source population, in addition
to possibilities as a cosmological probe and a method to determine
foreground galaxy bias. For now, we strongly suggest that 
there is adequate evidence for a weak lensing contribution to the
observed unresolved XRB - foreground galaxy cross-correlation.

Here, we have presented a hypothesis for the observed
cross-correlation between the unresolved XRB and foreground bright
galaxies using a
population of high redshift X-ray sources. The upcoming surveys with
CXO and XMM will allow the detection of such high redshift low
luminosity sources, as discussed in Haiman \& Loeb (1999) for the case
of CXO. The followup observations of such deep and planned X-ray
imaging
of wide fields will eventually test the presence of such a
population. In fact, the planned Guaranteed Time Observations (GTO) of
several deep fields with CXO, such as the Hubble Deep Field (HDF;
Williams et al. 1996), can easily be used to test the hypothesis
whether remaining contributors to the unresolved XRB are a low
redshift or high redshift population. The possibility that whether
the cross-correlation is due to clustering of low redshift sources or
lensing of high redshift sources can then be statistically studied
based on the observed redshift distribution and luminosity function of
X-ray emitting sources.

\begin{acknowledgements}

We acknowledge useful discussions and correspondences with Omar
Almaini, Zoltan Haiman and Lloyd Knox. 
Omar Almaini is also thanked for communicating
results from his analysis on the XRB-Galaxy cross-correlation 
and for answering various questions on the nature of XRB in general.
We also thank an anonymous referee for helpful comments and
suggestions on the manuscript and acknowledge partial support from a
McCormick Fellowship at University of Chicago.

\end{acknowledgements}


\begin{thebibliography}{}

\bibitem{} Almaini O., Shanks T., Griffiths R. E., et al. 1997,
MNRAS 291, 372

\bibitem[]{} Almaini O., Lawrence A., Boyle B. J. 1999, MNRAS in press
(astro-ph/9903178)
 
\bibitem{} Bartelmann M. 1995, A\&A 298, 661

\bibitem{} Carrera F.J., Barcons X., Butcher J. A., et al. 1995, MNRAS 275, 22

\bibitem{} Cen R., Ostriker J. P. 1999, ApJ 517, 31

\bibitem{} Cole S. 1991, ApJ 367, 45

\bibitem{} Cooray A. R. 1999a, New Astronomy (in press)

\bibitem{} Cooray A. R. 1999b, A\&A in press (astro-ph/9904246)

\bibitem{} Dolag K., Bartelmann M. 1997, MNRAS 291, 446

\bibitem{} Fiore F., La Franca F., Giommi P., et al. 
	1999, MNRAS in press (astro-ph/9903447)

\bibitem{} Haiman Z., Loeb A. 1999, ApJ submitted (astro-ph/9904340)

\bibitem{} Hasinger G., Turner J.T., George I.M., Boese G., 1992, GSFC
Calibration Memo CAL/ROS/92-001

\bibitem{} Hasinger G., 1999, in ``After the Dark Ages: When Galaxies
were Young'', Holt S. S., Smith E. P. (eds). AIP Press (Woodbury,
New York)

\bibitem{} Holland W. S., Greaves J. S., Zuckerman B., 
	et al. 1998, Nat 392, 788

\bibitem{} Hu W., Eisenstein D. J. 1998, ApJ 498, 497

\bibitem{}  Jain B., Seljak U. 1997, ApJ 484, 560

\bibitem{} Kaiser N. 1998, ApJ 498, 26

\bibitem{} Kauffmann G., White S. 1993, MNRAS 261, 921

\bibitem{} Lahav O., Fabian A. C., Barcons X., et al. 1993, Nat 364, 693

\bibitem{} Mellier Y., 1998, ARA\&A in press (astro-ph/9812172)

\bibitem{} Miyaji T. et al. 1994 ApJ 393, 134

\bibitem[]{} Miyaji T., Hasinger G., Schmidt M. 1998a, Proceedings of ``High
lights in X-ray Astronomy'', astro-ph/9809398

\bibitem[]{} Miyaji T., Ishisaki Y., Ogasaka Y., et al. 1998b, A\&A 334, L13

\bibitem{} Moessner R., Jain B. 1998, MNRAS 294, L18

\bibitem{} Peacock J. A., Dodds S. J. 1996, MNRAS 267, 1020

\bibitem{} Piro L., Scarsi L., Butler R. C., 1995, Proc. SPIE, 2517, 169

\bibitem{} Refregier A., Helfand J.D. McMahon R.G.,  1997, ApJ 477, 58

\bibitem{} Roche N., Griffiths R. E., Della Ceca R., et al. 1996, MNRAS 282, 820

\bibitem{} Sanz J. L., Mart\'inez-Gonz\'alez E., Ben\'itez N. 1997,
MNRAS 291, 418

\bibitem{}  Schneider P., van Waerbeke L., Jain B., Kruse G.,  1998, MNRAS 296, 873

\bibitem{} Schmidt M., Schneider D. P., Gunn, J. E. 1995, AJ 110, 68

\bibitem{} Smail I., Ivision R. J., Blain A. W., Kneib J.-P., 1999, 
in ``After the Dark Ages: When Galaxies were Young.'' eds. S. S. Holt
\& E. P. Smith (eds.)  (Woodbury: New York) astro-ph/9810281

\bibitem{} Soltan A. M., Hasinger G., Egger R., Snowden S. Tr\"umper
J. 1997, A\&A 320, 705

\bibitem{} Somerville R., Kolatt T., 1998, MNRAS in press (astro-ph/9711080)

\bibitem{} Viana P. T. P., Liddle A. R. 1996, MNRAS 281, 323

\bibitem{} Williams R. E., Blacker B., Dickinson M., et al. 1996, AJ 112, 1335



\end{thebibliography}
\end{document}